\documentclass[12pt]{iopart}

\def\met{\mbox{${\hbox{$E$\kern-0.6em\lower-.1ex\hbox{/}}}_T$ }} 
\def\stop{\widetilde{t}}
\def\sbottom{\widetilde{b}}
\def\squark{\widetilde{q}}
\def\sneutrino{\widetilde{\nu}}

\def\schi{\widetilde{\chi}}
\def\geqsim
{\mathrel{\rlap{\raise 0.53ex \hbox{$>$}}{\lower 0.53ex \hbox{$\sim$}}}}
\def\leqsim
{\mathrel{\rlap{\raise 0.4ex \hbox{$<$}}{\lower 0.72ex \hbox{$\sim$}}}}

\usepackage{epsf}

\begin{document}

\begin{flushright}
FERMILAB-Conf-00/001
\end{flushright}

\title[Physics Beyond the Standard Model]{Physics Beyond the
Standard Model
\footnote[1]{presented at the 1999 UK Phenomenology
Workshop, Durham; to appear in Journal of Physics G.}
}

\author{John Womersley\footnote[3]{e-mail: womersley@fnal.gov}}

\address{Fermi National Accelerator Laboratory,
P.O. Box 500, Batavia, IL 60510, U.S.A.}

\begin{abstract}
I shall briefly summarize the prospects for extending our
understanding of physics beyond the standard model within the
next five years.
\end{abstract}




In this necessarily brief presentation, I shall attempt to outline
how we may be able to discover physics beyond the standard model within
the next five years.  I shall interpret ``beyond the standard model'' 
to mean the physics of electroweak symmetry breaking, including
the standard model Higgs boson.  The nature of this TeV-scale
new physics is perhaps the most crucial question facing high-energy physics,
but we should recall (neutrino 
oscillations!) that there is ample evidence for interesting physics
in the flavour sector too.   In the next five years, before the LHC
starts operations, the facilities available will be 
LEP2, HERA and the Fermilab Tevatron.
I shall devote a bit more time to the Tevatron
as it is a new initiative for United Kingdom institutions.
The Tevatron schedule now calls for data taking in Run~II, using two
upgraded detectors, to begin on March 1,
2001, with 2~fb$^{-1}$ accumulated in the first two years.  A nine-month
shutdown will follow, to allow new silicon detector layers to be
installed, and then running will resume with a goal of accumulating
15~fb$^{-1}$ (or more) by 2006.

Where does the standard model stand, circa 2000?  We know that it
works very well, indeed some precision observables test it at the
10$^{-3}$ level\cite{sm}.  
All observations are consistent with a single, light
standard model (SM) Higgs boson, though no such beast has yet been observed.
As of Autumn 1999\cite{lepc},
Higgs masses less than 106~GeV are excluded by direct searches and
greater than 245~GeV are excluded by fits to precision SM 
observables\cite{gross}.   
Despite this consistency with the SM, there are strong general 
arguments for the existence
of new physics at the electroweak scale (250~GeV--1~TeV).  The goodness of
the SM fits may perhaps suggest that this new physics is weakly coupled.  
In addition 
there are indirect pointers that may suggest supersymmetry (SUSY) is a prime 
candidate for the new physics, though once again, all direct searches for
supersymmetric particles have so far proved negative. Data from
LEP2 have been used to rule out\cite{gross} squarks (stop
and sbottom) below 80--90~GeV, sleptons (selectron,
smuon and stau) below 70--90~GeV, charginos below 70--90~GeV, and
the lightest neutralino below 36~GeV.  Searches for squarks and
gluinos, stop, sbottom, charginos and neutralinos at the Tevatron
have also all proved negative and yield comparable limits\cite{aurore}.

Your mission, should you choose to accept it, is then clear.  At
your earliest convenience, please carry out one (or more!) of the
following challenges:
\begin{itemize}
\item discover the standard model Higgs;
\item discover or exclude the lightest Higgs of minimal SUSY (with masses up
to $m_h \sim 130$~GeV);
\item discover one or more superpartners;
\item exclude (or at least disfavour) 
supersymmetry at the TeV scale by discovering some other new physics.
\end{itemize}
We know the LHC can address all of these questions given sufficient
luminosity.
But can any of this be done sooner, {\it i.e.} in the next five years? 

 
\section{The Standard Model Higgs}

The search for the SM Higgs at LEP2 is well understood. The reach
depends on the centre of mass energy which can be achieved, and the
integrated luminosity obtained.  For 150~pb$^{-1}$ per experiment at
$\sqrt{s}=200$~GeV, Higgs masses $m_H < 109$~GeV will be ruled 
out\cite{gross}.
The LEP energy may be pushed a few GeV higher, but to go much
beyond this will require the use of the Tevatron.  Much interest in 
the Tevatron's potential for Higgs searches was sparked by 
work carried out during the Run~II SUSY/Higgs workshop in 
1998\cite{shw}.

\begin{figure}[t]
\begin{center}
\begin{tabular}{cc}
\epsfysize=6cm
\epsfbox{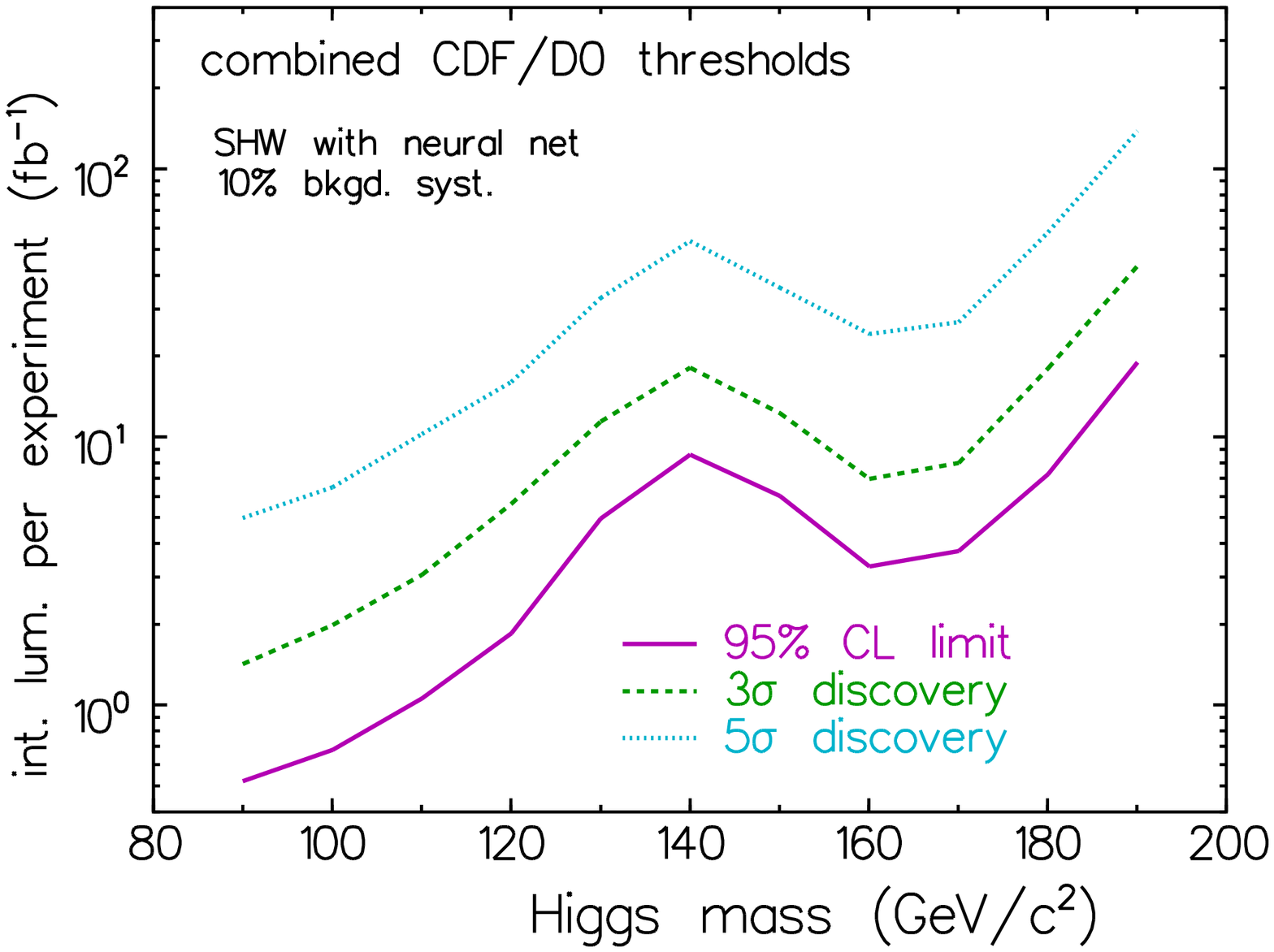}&
\epsfysize=6.7cm
\epsfbox{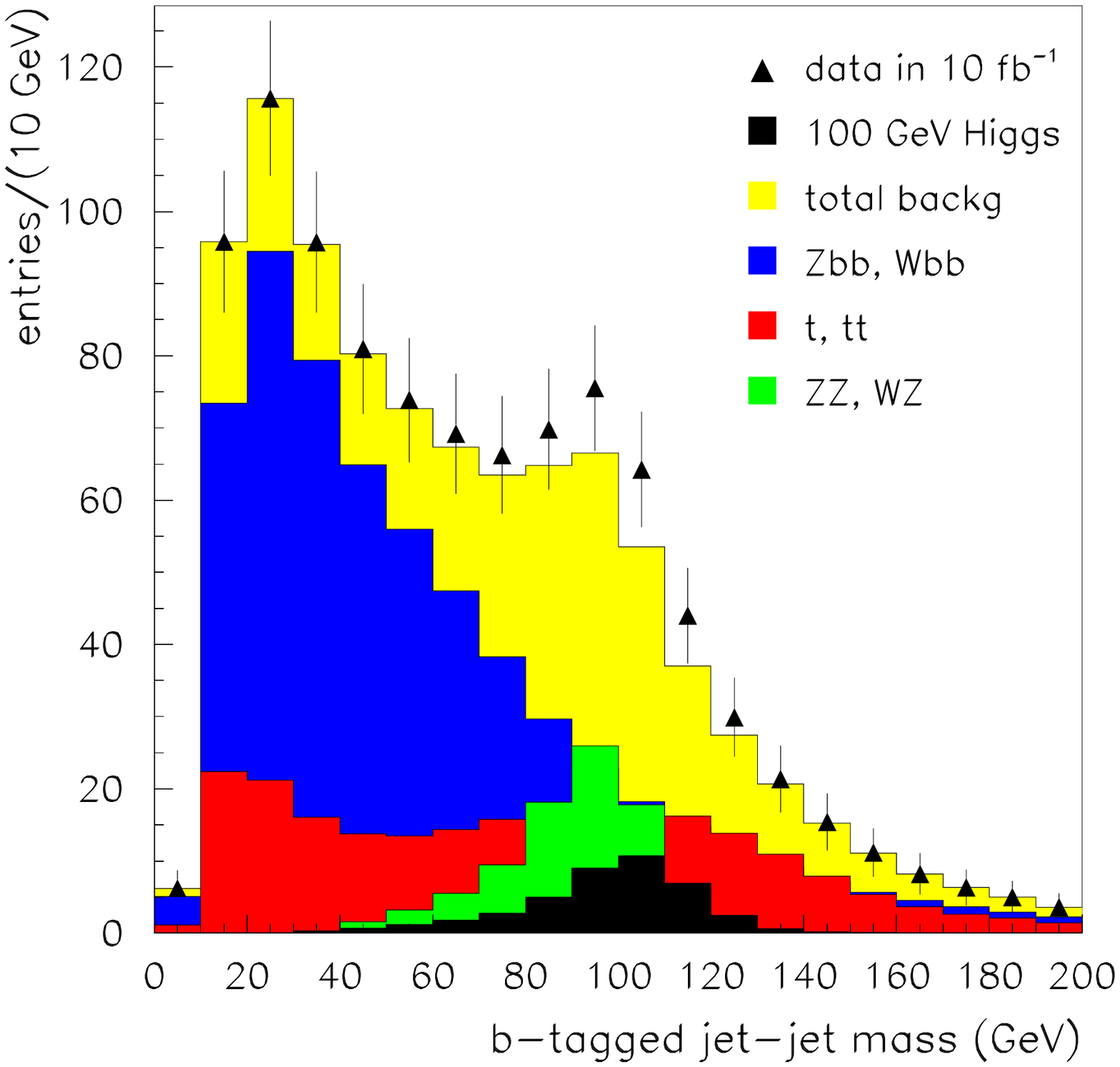}\\
\end{tabular}
\end{center}
\caption{(left) Luminosity required at the Tevatron for Standard Model
Higgs searches as a function of $m_H$. Background uncertainties of
10\% have been included.  
(right) Higgs signal ($m_H = 100$~GeV) and background in the channel
$p\overline p \to ZH \to \nu\overline\nu b\overline b$,
for 10~fb$^{-1}$.}
\label{fig:higgs}
\end{figure}

The Higgs cross section at the Tevatron is large (1~pb for $m_H \sim 100$~GeV)
but since the $gg \to H \to b \overline b$ process dominates,  there
is a huge QCD background.  For Higgs masses below 130--140~GeV,
the best potential seems to come from the
processes where a $W$ or $Z$ is produced in association with the Higgs:
\begin{itemize}
\item 
$WH \to \ell\nu b\overline b$:
Backgrounds from $W b\overline b$, $WZ$, $t\overline t$, single top.
A factor of 1.3 improvement in signal to background has been demonstrated in
this channel by using a neural network compared to standard cuts.  It is
also possible that additional gains may be had if the angular distributions
($WH$, spin zero, vs. $Wb\overline b$, spin one) can be exploited.
$WH \to q \overline q^\prime b\overline b$ is overwhelmed by the 
QCD background.
\item
$ZH \to \ell\ell b\overline b$:
backgrounds from $Z b\overline b$, $ZZ$, $t\overline t$.
\item
$ZH \to \nu\nu b\overline b$:
backgrounds from QCD, $Z b\overline b$, $ZZ$, $t\overline t$.
This requires a relatively soft missing $E_T$ trigger (35~GeV) but
is powerful because of the large $Z \to \nu\nu$ branching ratio.
\end{itemize}
For Higgs masses above 130--140~GeV:
\begin{itemize}
\item
$gg \to H \to WW^*$:
backgrounds from Drell-Yan, $WW$, $WZ$, $ZZ$, $t\overline t$, $Wt$.
The initial signal to background ratio is 1:140, so many, rather finely
tuned, selections are
required, culminating in angular cuts to separate the signal from the
``irreducible'' $WW$ background.
\end{itemize}
Combining all these search channels, and assuming a 10\% systematic
error on the backgrounds, one finds the sensitivity shown 
in Fig.~\ref{fig:higgs}.
With 2~fb$^{-1}$ of data, only a modest extension of the LEP reach
is possible. With 15~fb$^{-1}$, on the other hand, it should be possible to
either exclude, or observe at the 3--5$\sigma$ level, 
a SM Higgs up to $m_H \sim 180$~GeV.

This is an exciting prospect. Is it credible?  In my view, yes: it is an 
exercise similar in scale to the top discovery, with a similar number of
backgrounds, and requiring a similar level of detector understanding. 
It will be harder --- the irredicible signal to background 
ratio is worse --- but it has caught the imagination of the experimenters.
One problem with the studies so far (in my opinion) is the
$b \overline b$ mass resolution.  Can the assumed resolution really
be attained in a high luminosity environment?  The resolution not
only affects the 
mass bins over which a Higgs signal would be smeared,
but a detailed understanding of the shape of the $b \overline b$ mass
spectrum will be needed to separate a putative Higgs signal from
the nearby
mass peak due to $Z \to b \overline b$ decays (see Fig.~\ref{fig:higgs}).  
For this reason, calibration samples of
$Z \to b \overline b$ events will be very important. Tens of thousands 
of events (after cuts) are expected
for 2~fb$^{-1}$, using a displaced vertex trigger, in both CDF and D\O.

\section{Minimal Supersymmetry}

Even the minimal spectrum of supersymmetry (the SM particles plus two Higgs
doublets and their superpartners) can have many faces.  Firstly,
whether or not $R$ parity
is conserved determines whether the lightest supersymmetric particle 
(LSP) will or will not be stable.   
Secondly, how supersymmetry is broken determines the mass heirarchy of
states and hence their decays.  The typical benchmark, 
the supergravity-inspired or ``minimal SUGRA'' model, is determined by five
parameters $m_0$, $m_{1/2}$, $A_0$, $\tan\beta$ and ${\rm sign}\ \mu$.
In this picture, radiative electroweak symmetry breaking occurs
naturally from the large top mass.  The lightest neutralino is the LSP;
the two lightest neutralinos, the lighest chargino and the lightest
Higgs $h$ are all ``light'', while the other charginos and neutralinos,
squarks, gluinos and the other Higgs states are all ``heavy'' (above a
few hundred GeV, perhaps).  In gauge-mediated models, by contrast,
the LSP is a gravitino, and there are signatures with photons and/or
slow moving particles which may decay within or outside the detector.
Recent anomaly-mediated models suggest that the lightest chargino and
neutralino may be almost degenerate.

At LEP2, the increased centre of mass energy will allow the present
SUSY limits to be raised by 5--10~GeV (my guess) with full use of 
data at and above 200~GeV. 

At the Tevatron, the highest cross sections are for the pair
production of colored particles like squarks and gluinos. As long as
$R$-parity is conserved, the signature will be jets plus missing
transverse energy ($\met$).  The Run~I D\O\ analysis\cite{d0susy} 
required  three jets (one with $E_T > 115$~GeV) and $\met > 75$~GeV;
it excludes squark masses below about 250~GeV and
gluino masses below about 200~GeV. 
This search may be combined with complementary channels where one or more 
leptons are required, together with jets and $\met$\cite{d0ll}.
With 2~fb$^{-1}$ of data the reach will 
approximately be doubled, to gluino masses of order
400~GeV\cite{r2susy}.  
For masses much larger than this, falling parton distributions
kill the production rate very quickly.  To extend the reach in 
parameter space, chargino/neutralino production will become increasingly
important.  Present searches in the ``golden'' trilepton mode
do not really constrain models, but with 2~fb$^{-1}$ of data, 
chargino masses up to 180~GeV should be probed 
(150~GeV if $\tan\beta$ is large)\cite{r2susy}.  
This places an emphasis on
low-$p_T$ lepton triggering, since one or more of the leptons tends to
be soft.  It would also help a lot if $\tau$ modes could be included.

In many models, the stop and sbottom squarks are significantly lighter
than the others.  It is therefore interesting to search for them
separately. The decay channels involve $b$ or $c$ jets, and for
stop there is also the possibility of $t \to \stop$ decays 
or vice versa (depending on the masses).  Present CDF 
limits\cite{cdfstop} explore $m_{\stop} \leqsim 120$~GeV and 
$m_{\sbottom} \leqsim 145$~GeV; with 2~fb$^{-1}$, the sensitivity should
improve to about 200~GeV\cite{demina}.

Much interest in gauge-mediated supersymmetry was sparked a few
years ago by the observation of a single $ee\gamma\gamma\met$ 	
event at CDF\cite{gamgam}.  
This final state is consistent with selectron
production in a gauge mediated scenario.  All we can now say is that
searches for the expected related signatures have all proved negative:
$\gamma\gamma\met$ and $\gamma +{\rm jets}+\met$ at the Tevatron\cite{aurore}, 
and one or more photons plus missing energy at LEP\cite{lepgmsb}.

This type of signature is just one example of what may be expected in
gauge mediated models.  Depending on the details of the model,
the next-to-lightest superpartner or NLSP
(which then decays into the gravitino LSP)
may be a neutralino, a stau, or effectively more than one slepton 
state (if they are almost degenerate in mass).  Prompt decays of the NLSP will
then give final states containing photons$+\met$, taus, and multileptons,
respectively.  Searching for such signatures is relatively 
straightforward.  A more challenging possibility is that the NLSP
has a finite decay length --- since this is determined by the SUSY breaking 
interactions it is totally unknown.   For neutralino NLSP's, decay
lengths $c\tau$ more than a few metres will give standard $\met$ signatures
since the NLSP will escape the detector.
Decay lengths of a few centimetres to a metre result in photons which do not
point back to the primary vertex: searches at LEP\cite{lepgmsb} 
for such photons
have excluded neutralino NLSP masses less than 85~GeV, for $c\tau < 1$~m.
In Run~II at the Tevatron, D\O\ will be able to make use of
new preshower detectors upstream of the 
electromagnetic calorimeter to detect non-pointing
photons.  The resolution in the distance of closest approach to the vertex
will be 2.2~cm along the beamline and 1.4~cm radially\cite{longlived}.

Charged NLSP's with long decay lengths appear as massive, slow-moving
particles which exit the detector (so called ``cannonballs''). LEP
limits exclude stau NLSP masses less than 76~GeV, or slepton co-NLSP
masses less than 85~GeV\cite{gross}. 
In Run~II at the Tevatron, $dE/dx$ and timing
information will be available (using time of flight counters in CDF, and the
muon system in D\O).  CDF expect sensitivity up to stau masses of
180~GeV using the time of flight system\cite{r2gmsb}.  Short decay lengths 
$c\tau \leqsim 1$~cm will give reconstructable impact parameters in the
vertex detectors; 1~cm~$\leqsim c\tau \leqsim 1$~m is harder, especially to
trigger.  (With enough data, a combination of impact parameter and 
cannonball searches may be sensitive enough to exclude this intermediate 
region as well.) A general problem with understanding 
delayed decays is that event generators
are not widely available, and the interface to the detector 
simulation is non-trivial.

Recently there has been considerable interest in anomaly-mediated SUSY
models.  The phenomenology of such models contains a light chargino
which is almost degenerate with the LSP (a neutralino).  Such small
mass differences may result in delayed chargino decays, with
cannonball type signatures, or decays inside the detector 
${\schi^{\pm}} \to {\schi^0} + {\rm soft}\ \pi$: very challenging indeed.  

There is also considerable interest in scenarios with large extra
dimensions.  Gravitons may propagate into the higher dimensional
space.  Searches have been carried out\cite{gross} for 
$e^+ e^- \to \gamma+$nothing, 
$p\overline p \to \gamma+$nothing, and 
$p\overline p \to$jet$+$nothing. Indirect effects have been searched 
for in $e^+ e^- \to \gamma\gamma,\ \mu^+\mu^-,\ \tau^+\tau^-$.

Turning now to SUSY with $R$-parity violating decays,
the usual assumption is that the $R$ violating couplings
are small compared with the $R$ conserving ones. The
production processes and subsequent
decay chain then proceed as in minimal SUGRA, but the final LSP's decay
via baryon or lepton number violating interactions.  Hence there
is no $\met$ from the escaping LSP, but there are usually
multi-leptons from cascade decays and $\met$ from neutrinos.
Both at LEP\cite{gross} 
and the Tevatron\cite{d0rpv}, it is found (perhaps surprisingly)
that the excluded regions end up very comparable to those for
mSUGRA with $R$ conserved.
$R$-violating couplings could also enter in the production of 
SUSY particles. At HERA, ``leptoquark'' searches have been 
interpreted in terms of $ep \to \squark$\cite{hera}, 
while at LEP, limits have been placed on processes like 
$e^+e^- \to \sneutrino \to \tau^+\tau^-$\cite{leprpv}. 

Both at LEP2 and the Tevatron, we therefore
have a basic menu of supersymmetry
searches which is well-defined and we should have no trouble
in exploring:
\begin{itemize}
\item
minimal SUGRA;
\item
gauge mediated supersymmetry with prompt photon signatures;
\item
some subset of $R$ violation.
\end{itemize}
Our concerns are what we may have forgotten, especially at 
the Tevatron where triggering is a crucial issue.  Can we cover:
\begin{itemize}
\item
slow moving massive particles;
\item
gauge mediated supersymmetry with detached photons or taus;
\item
anomaly-mediated supersymmetry ({\it e.g.} $\schi^{\pm} \to \schi^0 +{\rm soft}\ \pi$);
\item
general extra dimension signatures, {\it etc?}
\end{itemize}
A critical look at the proposed trigger lists is probably
called for. 

\section{Supersymmetric Higgs}

At LEP2, two complementary processes allow the neutral Higgs states of
supersymmetric models to be explored:  $e^+e^- \to (h/H)Z$ and
$e^+e^- \to (h/H)A$.  Combining these channels and the four experiments,
one can exclude (Summer 1999\cite{gross}) $m_h < 81$~GeV and  $m_A < 81$~GeV.
Moreover, $0.9 < \tan\beta < 1.6$ ($0.6 < \tan\beta < 2.6$) is
excluded for maximal (and for minimal) stop mixing.  The $\tan\beta$
exclusion is very sensitive to the top mass, and it should be noted
that there is no excluded range in  $\tan\beta$ if $m_t = 180$~GeV.
General scans of minimal SUSY parameter space find some points
with strange decay patterns that can evade the limits, but these
are rather few.  Invisible Higgs decays have been included in the
searches.

In Run~II of the Tevatron, rather stringent limits will be set on
the SUSY Higgs sector, given sufficient luminosity, since the 
whole allowed mass range for the lightest Higgs $h$ ($m_h \leqsim 130$~GeV) 
is covered.  One may also exploit the enhanced couplings between
the Higgs states and $b$ quarks at large $\tan\beta$ to search
for $p\overline p \to b\overline b (h/A) \to 4b$.  Using present
data and requiring three $b$-tagged jets, CDF have been able
to explore the region of $\tan\beta \sim 50$ and above\cite{valls}.  
With 10~fb$^{-1}$ the sensitivity will extend to $\tan\beta = 30$ for
$m_A$ up to 150~GeV.

Supersymmetry also predicts the existence of charged Higgs states.
Searches for pair production at LEP have excluded masses 
$m_{H^\pm} < 77$~GeV\cite{gross}.  
In minimal SUSY models, it is expected
that $m_{H^\pm} > m_W$, and LEP is not really sensitive to this
region.  Searches at the Tevatron have concentrated on the
production of $H^\pm$ in top decays\cite{chiggs} in competition with the
standard $t\to W$ mode. This has been carried out 
both as an ``appearance'' experiment (looking for
$H^\pm \to \tau$) and as a ``disappearance'' experiment
(looking for fewer than expected $t \to e\ {\rm and\ }\mu$).
Present limits are sensitive only for $\tan\beta < 1$ and $>40$,
but with 2~fb$^{-1}$ the sensitivity will extend 
to $\tan\beta \sim 2$ and 20. 

\section{Non-supersymmetric Electroweak Symmetry Breaking}

There is no fully worked-out scenario for electroweak symmetry
breaking through a new strong interaction, but this does not mean
that it is not possible.  Indeed, many schemes have been outlined
in some detail, and a straw-man technicolor model is now
implemented in {\sc pythia}\cite{pythia}.  In general, dynamical symmetry
breaking schemes like technicolor and topcolor predict:
\begin{itemize}
\item
new particles in the mass range 100~GeV -- 1~TeV,
\item
with strong couplings and large cross sections,
\item
decaying to vector bosons and (preferentially 
third generation?) fermions.
\end{itemize}
Recently L3 has reported searches for technicolor resonances
at LEP2\cite{l3tc}, and several technicolor and topcolor searches 
have been made on present Fermilab data. All are negative so far.
With Run~II, the reach of the Tevatron will be greatly
extended.

Besides all of the above scenarios, one should always be looking
(at LEP, HERA and the Tevatron) for anything unexpected:
\begin{itemize}
\item leptoquarks
\item fourth generation fermions, or isosinglet fermions
\item $W^\prime$ and $Z^\prime$
\item contact interactions or compositeness, {\it etc.}
\end{itemize}

\section{Conclusions}

There are plenty of opportunities for us to find ``something new'' before
the LHC starts operation.  The standard scenarios have been explored in
some detail, but different supersymmetry breaking schemes can produce
radically different signatures.  Theoretical fashion moves fast, and
event generators are not always available.  We need to keep an open
mind in our searches.  This is easy to say once one has the data in hand;
the challenge right now is to ensure that we do our best to trigger on all 
possible interesting things, especially in the hadron collider environment.

\section*{Acknowledgments}

I would like to note my appreciation for
the hospitality of St.~John's College, Durham,
and thank M.~Whalley, J.~Forshaw and 
E.W.N.~Glover for all their organisational work.

\end{document}